\documentstyle [twocolumn,prl,aps,epsfig] {revtex}

\begin{document}
\draft
\flushbottom
\begin{title}
{\bf Spontaneous symmetry breaking in single and molecular quantum dots}
\end{title} 
\author{Constantine Yannouleas and Uzi Landman} 
\address{
School of Physics, Georgia Institute of Technology,
Atlanta, Georgia 30332-0430}
\date{Phys. Rev. Lett. {\bf 82}, 5325 [28 June 1999]; received 30 December
1998}
\maketitle
\begin{abstract}
Classes of spontaneous symmetry breaking at zero and low magnetic fields
in single quantum dots (QD's) and quantum dot molecules (QDM's) are
discussed in relation to the ratio $R_W$ between the interelectron 
Coulomb repulsion and the harmonic confinement, using spin-and-Space 
unrestricted Hartree-Fock calculations. These include: Wigner crystallization 
for $R_W > 1$, and formation of non-crystallized electron
puddles localized on the individual dots in QDM's, as well as spin-density
waves in single QD's, for $R_W < 1$.
\end{abstract}
\narrowtext

Two-dimensional (2D) electron gases have provided (e.g., the fractional 
quantum Hall effect \cite{tsg,laug}), and continue to provide 
(e.g., a charge-density wave at higher Landau levels \cite{cdwl}) a source of 
discovery of remarkable many-body phenomena.
Recently, 2D artificial quantum dots (QD's) and quantum dot molecules
(QDM's) have become available, with the capability of 
controlling the dots' size, shape, and number $N$ of electrons 
\cite{taru,asho}. 

Single QD's are commonly referred to as ``artificial atoms'', 
since interpretations of transport and capacitance experiments
draw often on analogies between such artificial structures and
natural atoms \cite{taru,asho}. Underlying these analogies is an
effective (circular) central mean field (CMF) picture, with the electronic 
spectra exhibiting (at zero magnetic field) shell closures and
following Hund's rules for open shells. 
Indeed, in experiments on single QD's, the addition energy (AE) spectra
\cite{taru} exhibit maxima at the expected closed shells ($N=2$, 6, 12), and
at the mid-shells ($N=4$, 9, and 16) in accordance with Hund's rule.

Here, using the self-consistent spin-and-Space unrestricted 
Hartree-Fock (sS-UHF) \cite{smey,note11} method, we discuss, for zero and low
magnetic fields ($B$), three types of spontaneous symmetry breakings (SB) 
in circular single QD's and in lateral QDM's 
(i.e., formation of ground states of lower symmetry
than that of the confining potentials \cite{pald}). 
These include: (I) Wigner crystallization (WC) \cite{wign} in both QD's
and QDM's, i.e., (spatial) localization of individual electrons, (II) 
formation of electron puddles (EP's) in QDM's, that is localization of the
electrons on each of the individual dots comprising the QDM, but without
crystallization within each dot, and (III) pure
spin-density waves (SDW's) which are not accompanied by spatial localization
of the electrons \cite{kosk}).
Furthermore, we show that CMF descriptions at
zero and low magnetic fields may apply only for low values of the parameter
$R_W \equiv Q/\hbar \omega_0$, where $Q$ is the 
Coulomb interaction strength and $\hbar \omega_0$ is the parabolic
confinement; $Q=e^2/\kappa l_0$, with $\kappa$ being the 
dielectric constant, $l_0=(\hbar/m^* \omega_0)^{1/2}$ the spatial 
extension of the lowest state's wave function in the parabolic confinement,
and $m^*$ the effective electron mass. 
With the sS-UHF, we find that WC occurs (SB of type I) in both QD's and
QDM's for $R_W > 1$. For QDM's with $R_W < 1$, WC does
not develop and instead EP's may form (SB of type II). 
We note here that
while certain quantum-mechanical studies of electron localization
(WC at high $B$) in single QD's have been discussed previously
\cite{mk,mmk}, this is the first study to explore, using a self-consistent 
quantum-mechanical treatment, broken symmetry WC and EP states in the $B=0$ and
small $B$ regimes for both circular single QD's and lateral QDM's,
thus providing new insights into the nature of these systems.
Additionally, for single QD's with $R_W < 1$ and $N \leq 20$, we find
that in the majority of cases the ground states exhibit CMF 
behavior without symmetry breaking; however, in several instances
(e.g., $N=14$), a pure SDW (SB of type III) develops.

The many-body hamiltonian for our problem is 
${\cal H} = \sum_{i=1}^N H(i) + \sum_{i < j}^N e^2/
\kappa | {\bf r}_i -{\bf r}_j |~$,
where the second term corresponds to the interelectron Coulomb repulsion.
The single-particle hamiltonian 
$H (i) = H_0 (i) + V_{\text{neck}} (i) + H_B (i)$ 
contains a term describing the motion of an electron in a 2D
two-center-oscillator (TCO) \cite{note1} confinement, i.e.,
%\begin{equation}
$H_0 (i) =  { {\bf p}_i^2 }/{2m^*} + 
m^* \omega_0^2 ( x_i^2 + y_{t,i}^{\prime 2} )/2$,
%\label{h0}
%\end{equation}
where $y_{t,i}^\prime = y_i - \widetilde{y}_t$; $t=l$ for $y_i < 0$ (left)
and $t=r$ for $y_i > 0$ (right), with $\widetilde{y}_l < 0$ and
$\widetilde{y}_r > 0 $ being the centers of the left and right oscillator.
In the QDM, the dots are joint smoothly via a neck
described by the term $V_{\text{neck}}(i)$ (4th-order polynomial in
$y_{t,i}^\prime$ \cite{barn,note1}) 
which allows variations of the bare interdot barrier
height $(V_b)$ for any distance $d=\widetilde{y}_r - \widetilde{y}_l$;
for $d=0$, the system reduces to a single harmonically confined 
QD. Magnetic-field effects are included in 
$H_B(i) = [({\bf p}_i - e{\bf A}/c)^2 - {\bf p}_i^2 ]/2m^*
+ g^* \mu_B {\bf B} {\bf \cdot} {\bf S}_i/\hbar$,
where ${\bf A}_i = B(-y_i/2,x_i/2,0)$ and the last term is the
Zeeman interaction with an effective factor $g^*$, ${\bf S}_i$ is the
electron spin, and $\mu_B$ the Bohr magneton.
To solve the sS-UHF equations \cite{note50}, we use a (variable with $d$) 
basis consisting of the eigenstates of $H_0(i)$ (due to it's separability,
the eigenfunctions of $H_0(i)$ can be expressed as
products of 1D harmonic-oscillator wave functions in $x_i$ and parabolic
cylinder functions in $y_{t,i}^\prime$ \cite{note1}).
In all calculations, we used $m^*=0.067 m_e$ (GaAs) and $\hbar \omega_0=5$
meV. For $\kappa$, we used the GaAs value of 12.9 (i.e., $R_W=1.48$),
as well as $\kappa=20$ (i.e., $R_W=0.95$, corresponding to a weakened 
interelectronic repulsion, due e.g. to the effect of the finite thickness of 
the dots).

The spatial distributions of the electronic densities for 
QD's and QDM's reveal in almost all cases that for $R_W=1.48$ 
(i.e., $\kappa=12.9$) the ground-state solutions are Wigner crystallized. 
An example of such a (finite) Wigner crystal with $B=0$
is shown in Fig.\ 1(a) for a $d=70$ nm
($-\widetilde{y}_l=\widetilde{y}_r=35$ nm), $V_b=10$ meV closed-shell QDM 
with $N=12$. The WC is portrayed by 6 well-resolved humps (3 in each well) for
both the up ($\uparrow$) and down ($\downarrow$) spins, and by 6 humps and
6 troughs for the spin density ($\uparrow - \downarrow$);
note that the density peaks for the two spin
directions do not overlap. Formation of such ``Wigner supermolecules'' (WSM's)
in QDM's is analogous to that of ``Wigner molecules'' (WM's) \cite{mmk} in 
single QD's. The appearance of such a WC is a consequence of the large value 
of $R_W$ (1.48 in this case). The mean distance between neighboring  density 
maxima inside each of the coupled dots equals $\bar{r} \approx 20$ nm, i.e.,
roughly twice larger than the effective Bohr radius 
$a^*_B (\kappa=12.9) = \hbar^2 \kappa /m^* e^2 = 10.188$ nm. 
Inspection of the wave functions shows that this case 
corresponds to an intermediate electron-density regime, where spatial
localization of {\it individual\/} electrons emerges, 
but with finite-amplitude contributions of each of the wave 
functions to several of the density peaks (i.e., ``weak'' WM, see below);
full localization into a ``classical'' WC requires
even lower densities \cite{jaur}.

A magnetic field compresses the electronic orbitals
in the QDM and the consequent increase in Coulomb repulsion
promotes electrons to higher orbitals of larger spatial extension, with
an increase in the spin polarization (spin flip)
resulting in optimization of exchange-energy gain 
(for a description of such a process in single QD's, see Ref.\ \cite{asho}). 
An example is shown for the QDM in Fig.\ 1(b) for
$B=3$ T ($g^*=-0.44$), where two
of the down-spin electrons flipped, resulting in 8 up-spin and 4 down-spin 
electrons, accompanied by a reduced Wigner crystallinity (partial ``melting'')
of the WSM, portrayed by the less pronounced density peaks 
[compare Fig.\ 1(b) with Fig.\ 1(a)], and increased density
in the interdot region. 

Having discussed formation of a WSM (in a 12e closed-shell QDM) made
of Wigner molecules in each of the coupled dots, we display in Fig.\ 2
results, with $B=0$, for the ground state (singlet) of a closed-shell
($N=6$) single QD for two values of $\kappa$. For $\kappa=12.9$
(i.e., $R_W=1.48$), we observe again the emergence of a WM;
note in Fig.\ 2(a, bottom) the
six charge-density maxima arranged on a ring, with $\bar{r} \approx 20$ nm.
On the other hand, for the same single QD, but 
with a reduced Coulomb repulsion ($\kappa=20$, $R_W=0.95$), no WC
occurs; compare the charge densities in Fig.\ 2(b) and 2(a). 
Thus by varying $R_W$, one may cross the ``phase boundary'' 
separating the localized Wigner-crystallization and delocalized
(CMF) regimes. Furthermore, in the WC regime the electron
(charge)-localization is accompanied here by a SDW 
[see top panel in Fig.\ 2(a) and also in Fig.\ 1(a)].

The WM emerging for $R_W=1.48$ [Fig.\ 2(a)] is a weak one. The transition into
the ``strong''-WM regime, caused by an increase in the strength of the
Coulomb repulsion, is illustrated in Fig.\ 2(c) 
for the same QD but with $\kappa=6$,
i.e., $R_W=3.18$. The two WM isomers shown exhibit sharper electron density
peaks reflecting stronger localization (seen also from much reduced 
wave-function amplitudes at neighboring sites).
The geometries of the lower-energy fully spin-polarized, 
$P \equiv N\uparrow - N\downarrow=6$, isomer
[Fig.\ 2(c), right], and that of the higher-energy isomer with $P=0$
[Fig.\ 2(c), left] agree well with those determined classically 
(i.e., for $R_W \rightarrow \infty$) \cite{bolt}.

Consider next an open-shell $(N=6)$ QDM with $d=70$ nm, $V_b=10$ meV,
and $R_W=1.48$, for which CMF-type treatments (as well as  
local-spin-density functional, LSD \cite{barn}),
predict a total net spin polarization $P=2$
in accordance with Hund's rule, while we find here 
that the ground-state of the QDM is a Wigner-crystallized singlet, 
i.e. $P=0$ [see Fig.\ 3(a)], consisting of two
spin-polarized (triplet) WM's [formed inside the left and right dots; 
see Fig.\ 3(a), top]. 
An excited state of the molecule (with a 0.09 meV higher
energy) shown in Fig.\ 3(b) is also crystallized but 
with a net spin polarization
$P=2$; note the different spatial configurations of the
ground and excited states.
Reducing the $R_W$ value for the QDM to 0.95 (i.e., $\kappa=20$)
transforms the ground-state of the 6e QDM from the crystallized state
[Fig.\ 3(a)] into one consisting of
electron puddles [SB of type II, Fig.\ 3(c, left)]; 
here each of the EP's (on the left and right dots)
is spin-polarized with $P_l=1$, $P_r=-1$, and the singlet and triplet states
of the whole QDM are essentially degenerate. Note that the orbitals
on the left and right dots [see, e.g., those on the left dot in Fig.\ 
3(c, right)] are those expected from a CMF treatment,  
but with slight (elliptical) distortions
due to the interdot interaction. Only for much lower values of $R_W$
($\lesssim 0.20$, i.e., $\kappa \gtrsim 90.0$) Fermi liquid (delocalized)
bahavior is restored.

Results of sS-UHF calculations at $B=0$ for the AE's, 
$\Delta \varepsilon =E(N+1) - E(N) - [E(N)-E(N-1)]$, where $E(N)$ is the 
$N$-electron total energy, are shown for single QD's and QDM's in Fig. 4.
For the single QD's and for a rather wide range of 
$R_W > 1$ (e.g., see the curve marked $\kappa=12.9$), 
the AE's (corresponding mostly to WC states)
exhibit maxima at the same values of $N$ (see introductory paragraphs) 
as in the ``normal'' (i.e., $R_W < 1$ non-crystallized, CMF) regime.
However, while in the latter the spin-polarizations follow Hund's rule
(except for $N=14$, corresponding to a pure SDW state), those
in the WC states ($\kappa=12.9$) for $N \geq 8$ do not. 
For the QDM, a CMF treatment
predicts for sufficiently large interdot barriers (e.g., $V_b=10$ meV)
shell closures at $N=4$ and 12 (i.e., twice the single QD values) and (Hund's)
half-shell maximum spin polarizations (e.g., at $N=8$).
While shell-closure features are observed in the AE's of the QDM's with 
$\kappa=12.9$ (WC) and $\kappa=20$ [where EP's may form, see, e.g., Fig.\ 
3(c)] shown in Fig.\ 4 (more pronounced for $\kappa=20$), in both cases the 
spin polarizations do not in general follow Hund's rule (e.g., $N=6$ 
and 8).

In summary, using sS-UHF calculations, we discussed three classes of
spontaneous SB in QD's and QDM's at zero and low magnetic fields, i.e.,
formation of Wigner crystallized molecules and supermolecules for
$R_W > 1$, and non-crystallized electron puddles localized on the
individual dots in QDM's, as well as pure SDW's in single QD's (for $R_W <1$).
Further studies of such broken symmetries may include:
mapping of ``phase-boundaries'' through variations
of materials dependent (e.g., dielectric constant) and externally
controlled (e.g., gate voltages, interdot distances and barrier heights,
and magnetic fields) parameters, and probing of excitations 
and spin polarizations \cite{note20}. 

This research is supported by the US D.O.E. (Grant No.
FG05-86ER-45234).

\newpage
\begin{figure}
\caption{
sS-UHF results for a 12e QDM 
($d=70$ nm, $V_b=10$ meV, 
$\hbar \omega_0 = 5$ meV), without [in (a)] and with a magnetic field
$B=3$ T [in (b)]. For both cases, the bottom and middle panels 
correspond to the up-spin and down-spin electron distributions, respectively,
and the top ones correspond to the 
difference between them (spin density). 
Lengths ($x$ and $y$ axes) in nm, density distributions
(vertical axes) in 10$^{-3}$ nm$^{-2}$. $x$-axes, $y$-axes, and vertical-axes 
scales in (b) are the same as in (a).  
}
\end{figure}

\begin{figure}
\caption{
Total electronic density distributions [bottom of (a) and (b), and panel (c)], 
and spin density distributions [top of (a) and (b)] for a 6e single QD with 
$\hbar \omega_0=5$ meV and $B=0$; $\kappa = 12.9$ ($R_W=1.48$) in (a), 
$\kappa=20$ ($R_W=0.95$) in (b), and $\kappa=6.0$ ($R_W=3.18$) in (c).
The spin polarizations are $P=0$ in (a), (b), and (c, left), and $P=6$ for the
isomer in (c, right) whose energy is lower by 1.72 meV than that of the one
shown in (c, left). Units as in Fig.\ 1.
}
\end{figure}

\begin{figure}
\caption{
(a) Ground-state and (b) an excited-state Wigner-crystallized 
total electronic (charge)
density distributions (bottom panels) and spin-density distributions (top)
for a 6e QDM with $\hbar \omega_0=5$ meV, $d=70$ nm, $V_b=10$ meV,
$\kappa=12.9$, and $B=0$. (c, left): total electronic density for the same QDM,
but with $\kappa=20$, illustrating formation of (non-crystallized within the 
dots) EP's. (c, right): contour plots of the densities of the three individual 
orbitals localized on the left dot ($P_l=1$, with spin polarization of the 
orbitals as indicated). Units as in Fig.\ 1.
}
\end{figure}

\begin{figure}
\caption{
sS-UHF results for the addition energies ($\Delta \varepsilon$ vs. $N$)
of a single QD ($\hbar \omega_0=5$ meV, two upper curves) and for
a QDM ($\hbar \omega_0=5$ meV, $d=70$ nm, $V_b=10$ meV) calculated for
$\kappa=12.9$ and 20, and for $B=0$. Energies in meV. The spin polarizations,
$P= N\uparrow -N\downarrow$, are marked on the curves.
}
\end{figure}

\newpage
~~~~~~~~~\\
\newpage
\widetext
\begin{center}
{\large {\bf
{Erratum: Spontaneous symmetry breaking in single and molecular quantum dots
[Phys. Rev. Lett. 82, 5325 (1999)]}
}}\\
{Constantine Yannouleas and Uzi Landman}\\
%\address{
%School of Physics, Georgia Institute of Technology,
%Atlanta, Georgia 30332-0430 }

(Phys. Rev. Lett. {\bf 85}, 2220 [4 September 2000])
\end{center}
~~~~~~~~~~\\

In our recent Letter on spontaneous symmetry breaking in quantum dots (QD's),
we displayed in Fig.\ 4 addition energies $\Delta \varepsilon (N)=
E(N+1)-2E(N)+E(N-1)$, where $E(N)$ is the $N$-lectron ground-state total
energy, calculated with the spin-and-space unrestricted Hartree-Fock (sS-UHF)
method. Subsequent to the the publication of our Letter, we have performed
further systematic sS-UHF calculations, with the use of larger 
harmonic-oscillator bases and an implementation of an extensive search
for energy minima. While the behavior and magnitudes of $\Delta \varepsilon$
shown in Fig.\ 4 of our Letter are maintained, as well as our finding 
pertaining to the prevalent violation of Hund's first rule, our improved
calculations yield in certain instances different spin polarizations
$P \equiv N\uparrow -N\downarrow$, where $N\uparrow$ and $N\downarrow$
($N\uparrow + N\downarrow = N$) are the number of electrons with up and down
spins, respectively.

In Fig.\ 1 below, we display our new results for $\Delta \varepsilon$ in a
a single QD with GaAs parameters, i.e., $\kappa=12.9$, $\hbar \omega_0= 5$ 
meV, and $m^*=0.067 m_e$ (corresponding to the top curve in Fig.\ 4 of our 
Letter). The results of our calculations (solid 
dots) for the spin polarization given in the inset to Fig.\ 1 exhibit 
violation of Hund's rule (open squares) for $N=4,8,9,14,15,16,18$, and 22
(note that this violation appears already for $N=4$); however, as noted in
our Letter, the addition energies display maxima at closed shells
(i.e., at $N=6, 12$, and 20), as well as at the mid-shell closures
(i.e., at $N=4, 9$, and 16). In general, the non-Hund ground-state minima
are accompanied by energetically close spin isomers obeying Hund's rule,
and vice versa (e.g., for $N=17$ the energy difference between the Hund,
$P=3$, and the non-Hund, $P=1$, isomers is 0.05 meV).

%***************** begin figure 1 **************************
%\begin{minipage}[ht]{\textwidth}
\begin{figure}[ht]
\begin{center}
\epsfig{file=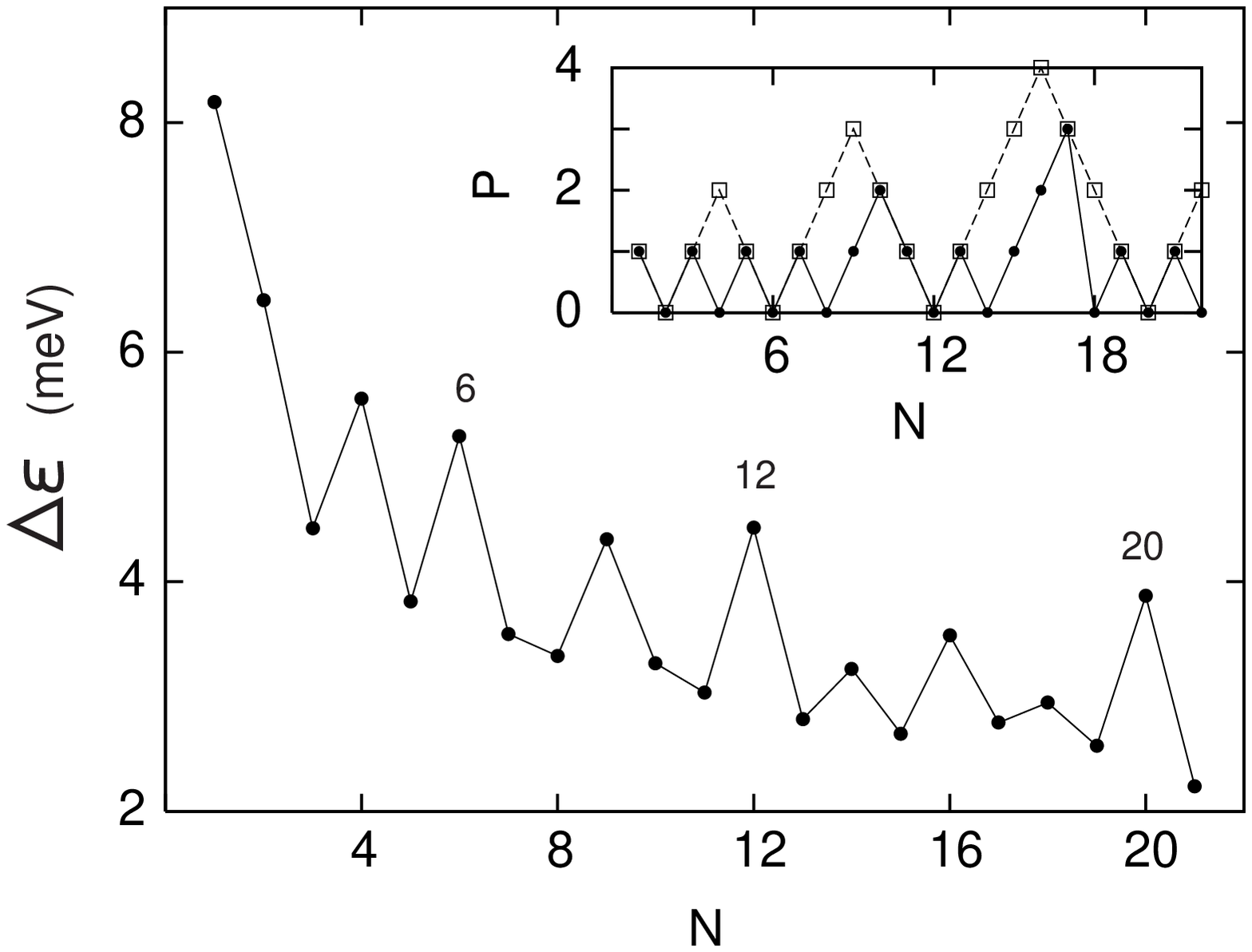,width=0.6\textwidth,clip=,angle=0}
\end{center}
%~~~~~~~~~~~\\
\caption{
sS-UHF results for the addition energies ($\Delta \varepsilon$ vs $N$) of a
single QD ($\hbar \omega_0=5$ meV, $\kappa=12.9$, $m^*=0.067 m_e$). The inset
displays the calculated (solid dots) spin polarizations, $P \equiv N\uparrow 
-N\downarrow$, as well as those (open squares) expected from Hund's first 
rule.
}
\end{figure}
%\end{minipage}
%***************** end figure 1 ************************** 
For the corresponding case of a quantum dot molecule (QDM) with
$\kappa=12.9$, $d=70$ nm, and $V_b=10$ meV (third curve from the top in 
Fig.\ 4 of our Letter), our improved calculation yields $P=0$ for $N=14$
(unlike the earlier value of $P=2$), with all the other spin polarizations 
remaining unchanged.

\end{document}